\documentclass[aps,prb,superscriptaddress]{revtex4}
\usepackage{graphicx,psfrag,bm,amsmath,amsfonts,amssymb,color,subfigure}

\newcommand{\be}{\begin{equation}}
\newcommand{\ee}{\end{equation}}
\newcommand{\beq}{\begin{eqnarray}}
\newcommand{\eeq}{\end{eqnarray}}
\newcommand{\ba}{\begin{array}}
\newcommand{\ea}{\end{array}}
\newcommand{\bea}{\begin{eqnarray}}
\newcommand{\eea}{\end{eqnarray}}
\newcommand{\ex}[1]{\mbox{e}^{#1}}

\newcommand{\eps}{\epsilon}

\newcommand{\mcal}[1]{\mathcal{#1}}

\newcommand{\bra}[1]{|#1\rangle}
\newcommand{\ket}[1]{\langle #1 |}
\newcommand{\braket}[2]{\langle #1 | #2\rangle}

\newcommand{\om}{\omega}
\newcommand{\Om}{\Omega}

\newcommand{\bma}{\begin{matrix}}
\newcommand{\ema}{\end{matrix}}

\newcommand{\elud}{e_{L\uparrow}^{\dagger}}

\newcommand{\eldd}{e_{L\downarrow}^{\dagger}}

\newcommand{\erud}{e_{R\uparrow}^{\dagger}}

\newcommand{\erdd}{e_{R\downarrow}^{\dagger}}

\newcommand{\hrud}{h_{R\Uparrow}^{\dagger}}

\newcommand{\hruds}{h_{R\sigma_h}^{\dagger}}
\newcommand{\ed}[1]{e_{#1}^{\dagger}}

\newcommand{\hd}[1]{h_{#1}^{\dagger}}

\begin{document}
\title{Coherent optical manipulation of triplet-singlet states in coupled quantum dots}
\author{Hakan E. T\"ureci}
\affiliation{Institute of Quantum Electronics, ETH-Z\"{u}rich,
CH-8093, Z\"{u}rich, Switzerland}
\author{J. M. Taylor}
\affiliation{Department of Physics, MIT, MA 02139, USA}
\author{A. Imamoglu}
\affiliation{Institute of Quantum Electronics, ETH-Z\"{u}rich,
CH-8093, Z\"{u}rich, Switzerland}
\date{\today}
\begin{abstract}
We show that spin-orbit coupling in a quantum dot molecule allows for coherent manipulation of two electron spin states using Raman transitions. Such two-electron spin states defined by the singlet and triplet states of two exchange coupled quantum dots can have favorable coherence properties. In addition, two of the four metastable ground states in this system can be used as {\sl auxiliary} states that could facilitate implementation of tasks such as mapping of spin states to that of a single propagating photon. We find that even weak spin-orbit effects --- manifesting themselves as slightly different $g$-factors for the electron and the hole --- would allow for the coherent Raman coupling of the singlet-triplet states. We also discuss the possibilities for implementing quantum optical techniques for spin preparation and manipulation.
\end{abstract}
\maketitle

\section{INTRODUCTION}

Over the past decade semiconductor quantum optical systems,
implemented in quantum wells and particularly quantum dots, have
been paradigmatic for the exploration of novel quantum mechanical
effects in the solid state~\cite{awschalom01}.
Potential applications as single photon
sources~\cite{MichlerKBSPZHI00,PeltonSVZSPY02} or as quantum bits
for quantum information storage and processing~\cite{imamoglu00}
have driven investigations of single quantum dots with ground states
containing zero or a single electron charge.  These systems bear marked
resemblance to noble gas and alkali atoms.  However, systems similar
to alkaline-earth atoms (with two valence electrons) or homo-nuclear
alkali molecules remain largely unexplored.  Such systems typically
have a more complex fine structure, leading to metastable spin
states with useful decoherence properties~\cite{haljan05} and have
well-established semiconductor realizations~\cite{krenner05,StinaffSBPKWDRG06}.

In this article we examine approaches for optically coupling the
four metastable ground (spin) states of a two-electron double
quantum dot system via optical Raman transitions.  These states,
split into a singlet and triplet manifold, have demonstrated useful
properties with respect to spin-related dephasing \cite{duan97,zanardi97,lidar98}, as seen in recent
experiments in electrically-controlled double quantum
dots~\cite{PettaJTLYLMHG05}.  Our approach for coupling singlet and
triplet states via optical fields can lead to the integration of optical manipulation,
measurement and entanglement techniques with demonstrated approaches
to controlling fine structure states in electrical quantum dots.

While we focus on the case of Zinc-Blende (III-V) semiconductor
quantum dots, we find that optical coupling of ground state spins
can be realized even with weak spin-orbit interaction, where a
sufficient condition is that the electron and hole states have
differing $g$-factors; this is believed to be the case for small
radius carbon nanotubes~\cite{Jarillo-HerreroSDKv04}. Thus our
approach for working with fine structure states can find wide
application in a variety of quantum dot systems.

\section{Coherent optical manipulation of coupled quantum dots
with strong-spin orbit effects}

In this section we develop an approach to coupling singlet and
triplet fine structure states of a double quantum dot system via
optical Raman transitions. We rely upon the dramatic difference
between exchange energies for two electrons on the same
quantum dot and two electrons in separated quantum dots to provide
a reliable means of using single spin selection rules to develop
controlled, two-spin selection rules.  We find that for
doubly-charged double quantum dots with spin-orbit coupling, such
as Zinc-Blende semiconductor quantum dots, polarization- and
energy-selective transitions between all fine structure states are
possible.  This allows techniques, such as STImulated Raman
Adiabatic Passage (STIRAP) to be used to initialize arbitrary
superpositions of singlet and triplet spin states.

\begin{figure}[hbt]
\includegraphics[clip,width=0.4\linewidth]{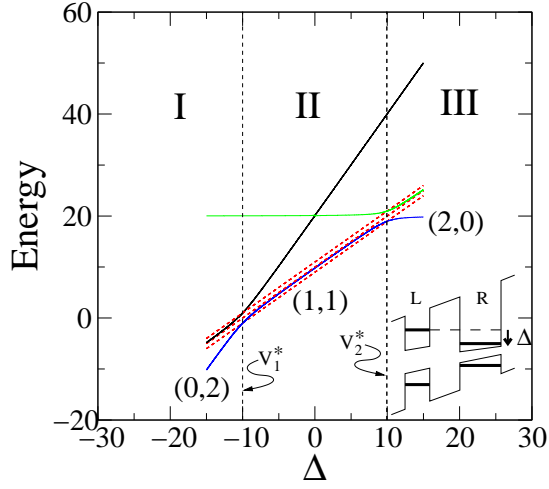}
\caption{The energy level structure of the coupled dot system as a
function of the detuning $\Delta$. The charge state of the ground
state follows the sequence $(0,2)$, $(1,1)$, $(2,0)$ from left to
right. $\Delta=V_i^*$ ($i=1,2$) denote the anticrossing points at
which the system undergoes a charge transition. The dashed lines
show the triplet states $\bra{(1,1)T_0}$ and $\bra{(1,1)T_\pm}$. The ground
state is always in the singlet configuration at low magnetic fields.
In the inset, we show the single-particle level structure of the
coupled quantum dots. The detuning $\Delta<0$ for the configuration
shown.} \label{figSTmblv}
\end{figure}

We consider here the optical transitions in a doubly-charged
coupled quantum dot (CQD), a situation that can for instance be
realized with an asymmetric pair of stacked InAs quantum dots
embedded in a Schottky diode structure \cite{StinaffSBPKWDRG06} in
an appropriate gate voltage regime. We assume that the left dot (L)
is blue-shifted with respect to the right dot (R) and that the
lowest conduction level of the right dot is detuned by $\Delta$ with
respect to the left one (see Fig.~\ref{figSTmblv}, inset). Higher
orbital states can be neglected as the associated transitions are
well separated in energy in the gate voltage regime of interest. For a range of gate voltages where the two
dots contain a total of two electrons, by fine tuning the voltage it
is possible to convert between atomic and molecular orbital states
\cite{StinaffSBPKWDRG06}, i.e., between charge states $(0,2)$,
$(2,0)$ and $(1,1)$. Labels $(m,n)$ here refer to the number of
electrons confined in the $({\rm left,right})$ dot. In the presence
of finite interdot tunneling and in the regime where $(1,1)$ is the
lowest energy charge configuration, the two resident electrons
hybridize resulting in an energetically isolated singlet-triplet
subspace where the effective degree of freedom is the total spin of
the two electrons. This regime is denoted by (II) in
Fig.~\ref{figSTmblv} and in what follows we will work in this
regime.

In the (1,1) regime (II), the ground state manifold is given by the
states $\bra{(1,1)S}  =  \frac{1}{\sqrt{2}}( \elud\erdd - \eldd\erud
) \bra{0}$, $\bra{(1,1)T_0}  =  \frac{1}{\sqrt{2}}( \elud\erdd +
\eldd\erud ) \bra{0}$, $\bra{(1,1)T_+}  =  \elud\erud  \bra{0}$ and
$\bra{(1,1)T_-}  = \eldd\erdd  \bra{0}$ with energies
$E_{S_0}=\Delta + E_c^{LR}-\mathcal{J}$, $E_{T_0,T_\pm}=\Delta +
E_c^{LR}$, where $\mathcal{J} \approx T_e^2/(E_C^{RR}-E_C^{LR})$.
The typical energies for an InAs self-assembled quantum dot  are
given by $E_C^{LL} \approx E_C^{RR} = V^{ee}_{LL,LL} \approx 20
\,\,\, meV$, $ E_C^{LR} = V^{ee}_{LL,RR} \approx 10 \,\,\, meV$,
where $V^{ab}_{ij,kl} = \int \int \, d\bm{r} \, d\bm{r}' \varphi^a_i
(\bm{r}) \varphi^a_j (\bm{r}) V_c(|\bm{r}-\bm{r}'|) \varphi^b_k
(\bm{r}') \varphi^b_l (\bm{r}')$ and $\varphi^a_j (\bm{r})$ are the
single-particle envelope wavefunctions on dot $j=L,R$ for conduction
band electrons ($a=e$) or holes ($a=h$). The tunneling matrix
element $T_e$ appearing in the exchange splitting $\mcal{J}$ is
given by $T_e=t_e + V^{ee}_{LL,LR} \simeq t_e + V^{ee}_{RR,LR}
\stackrel{<}{\sim} 1 \,\,\, meV$ and connects $\bra{(1,1)S}$ to
$\bra{(2,0)S}$ and $\bra{(0,2)S}$. Here, $t_e$ is the bare interdot
electron tunneling matrix element.

\begin{figure}[hbt]
\includegraphics[clip,width=0.2\linewidth]{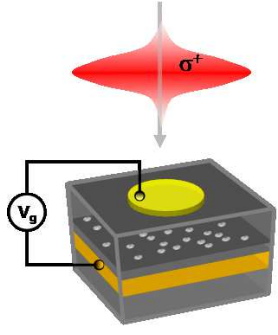}
\caption{Schematics showing the optical excitation of the CQD
system embedded in a Schottky-diode structure. The light pulse
$\Omega_+(t)$ is incident along the crystal growth direction
($z$-axis).} \label{figdotexschem}
\end{figure}

Consider now a right-hand circularly polarized ($\sigma_+$) optical
excitation with its optical axis along the heterostructure growth
direction ($z$-axis). This axis is defined with respect to the
crystal axes of the quantum well structure on which the dots are grown, which typically defines the axis
of shape asymmetry of the quantum dots (see
Fig.~\ref{figdotexschem}). The light-matter interaction Hamiltonian
in the dipole, rotating wave and envelope function approximations is
given by \be V_+ = \frac{i\hbar}{2} \Omega_+ \left( \mcal{M}_{RR}
\erdd\hrud + \mcal{M}_{LR} \eldd\hrud \right) \ex{-i\omega_+t} +
h.c. \ee Here, $\Omega_+ = \frac{1}{\sqrt{2}}
\ket{L^{e}=0,L_z^{e}=0\,} \, x+iy \, \bra{L^h=1,L_z^h=+1}$, where
$\bra{L^e=0,L_z^e=0\,}$ is the periodic part of the conduction band
Bloch wavefunction which has $s$-character, and
$\bra{L^h=1,L_z^h=+1}$ is that of the valence band electrons (or
holes) which has $p$-character, while $\mcal{M}_{RR}$ and $\mcal{M}_{LR}$ are the
overlaps of the electron and hole {\em envelope} wavefunctions
$_e\braket{R}{R}_h$ and $_e\braket{L}{R}_h$ respectively. The
implicit assumption here is that the light-hole levels and the
spin-orbit split-off hole-band is energetically well-separated from
the heavy-hole band denoted by
$\bra{\Uparrow}=\bra{J^h=3/2,J_z^h=3/2}$,
$\bra{\Downarrow}=\bra{J^h=3/2,J_z^h=-3/2}$
($\bm{J}^h=\bm{L}^h+\bm{S}^h$) so that their optical coupling can be
neglected, a well-justified approximation in Zinc-Blende (III-V)
semiconductor quantum dots \cite{Bahder90}. Note that it is the
correlation between the spin and spatial parts of the heavy-hole
states ( $\bra{\Uparrow}  = \bra{L_z^h=1;\uparrow}$,
$\bra{\Downarrow}  = \bra{L_z^h=-1;\downarrow}$) which enables us to
use selection rules based on pseudospin conservation.
The excited state manifold of doubly charged excitons ($X^{2-}$) is eight dimensional
\begin{align*}
& \bra{(2,1\uparrow)\sigma_h} = \elud\eldd\erud\hruds \bra{0} & E_1 = E^{LLR}\\
& \bra{(2,1\downarrow)\sigma_h} = \elud\eldd\erdd\hruds \bra{0} & E_2 = E^{LLR} \\
& \bra{(1\uparrow,2)\sigma_h} = \elud\erud\erdd\hruds \bra{0} & E_3 = E^{LRR}+\Delta \\
& \bra{(1\downarrow,2)\sigma_h} = \eldd\erud\erdd\hruds \bra{0} & E_4 = E^{LRR}+\Delta\\
\end{align*}
where $\sigma_h=\Uparrow,\Downarrow$. We assume that because of the
particular structure of the dots and the bias of choice, the
optically generated hole always resides on the right dot within its
lifetime. Here, $E^{LLR}=V^{ee}_{LL,LL} + 2V^{ee}_{LL,RR} -
V^{ee}_{LR,LR} - 2V^{eh}_{LL,RR} - V^{eh}_{RR,RR}$ for instance and
$E^{LLR}>E^{LRR}$ because of the e-h attraction. An additional
electron tunneling matrix element connects states
$\bra{(2,1\uparrow)\Uparrow}$ and $\bra{(1\uparrow,2)\Uparrow}$ as
well as $\bra{(2,1\downarrow)\Uparrow}$ and
$\bra{(1\downarrow,2)\Uparrow}$, which gives rise to an
anti-crossing at around the bias $\Delta \approx
E^{LLR}-E^{LRR}\approx 15-20  \,\,\, meV$. If we operate in the
(1,1) regime (II) close to $V=V_1^*$, we can safely neglect any mixing of these
$X^{2-}$ states.

Consider the action of $\sigma_+$ optical excitation on the $(1,1)$ ground state manifold
\begin{align*}
& V_+\bra{S} = -\mcal{M}_{RR}\bra{(1\downarrow,2)\Uparrow} - \mcal{M}_{LR} \bra{(2,1\downarrow)\Uparrow} \\
& V_+\bra{T_0} = -\mcal{M}_{RR}\bra{(1\downarrow,2)\Uparrow} + \mcal{M}_{LR} \bra{(2,1\downarrow)\Uparrow} \\
& V_+\bra{T_+} = \mcal{M}_{RR}\bra{(1\uparrow,2)\Uparrow} - \mcal{M}_{LR} \bra{(2,1\uparrow)\Uparrow} \\
& V_+\bra{T_-} = 0
\end{align*}
Here, we have discarded the common factor $(i\hbar/2)\, \Om_+$. Note
that there is no  optical transition from $\bra{T_-}$ under
$\sigma_+$ circular polarization (and similarly for $\sigma_-$ and
$T_+$ ). The following table illustrates the optical selection rules
for transitions from $\bra{(1,1)S}$ and $\bra{(1,1)T_0}$.
\begin{center}
\begin{tabular}{c|c|c|c}
& $\sigma_+$ & $\sigma_-$ & \\
\hline
$\mcal{M}_{RR}$ & $(1\downarrow,2)\Uparrow$ & $(1\uparrow,2)\Downarrow$ & $E_1$ \\
$\mcal{M}_{LR}$ & $(2,1\downarrow)\Uparrow$ & $(2,1\uparrow)\Downarrow$ & $E_3$
\end{tabular}
\end{center}
Since $\mcal{M}_{RR} \gg \mcal{M}_{LR}$, the upper row transitions are strongest, while the lower row may be neglected.

The e-h exchange interaction in the relevant optically excited
states would be negligibly small since the unpaired electron is
exchange coupled to the hole that resides in a different dot.
Starting with $\bra{(1,1)S,T_0}$, the states
$\bra{(1\downarrow,2)\Uparrow}$ and
$\bra{(1\uparrow,2)\Downarrow}$ will form the "bright" excitons
$X^{2-}$, whereas the states $\bra{(1\uparrow,2)\Uparrow}$ and
$\bra{(1\downarrow,2)\Downarrow}$ will form the
bright excitons for the $\bra{(1,1)T_\pm}$ subspace.

\begin{figure}[hbt]
\includegraphics[clip,width=0.7\linewidth]{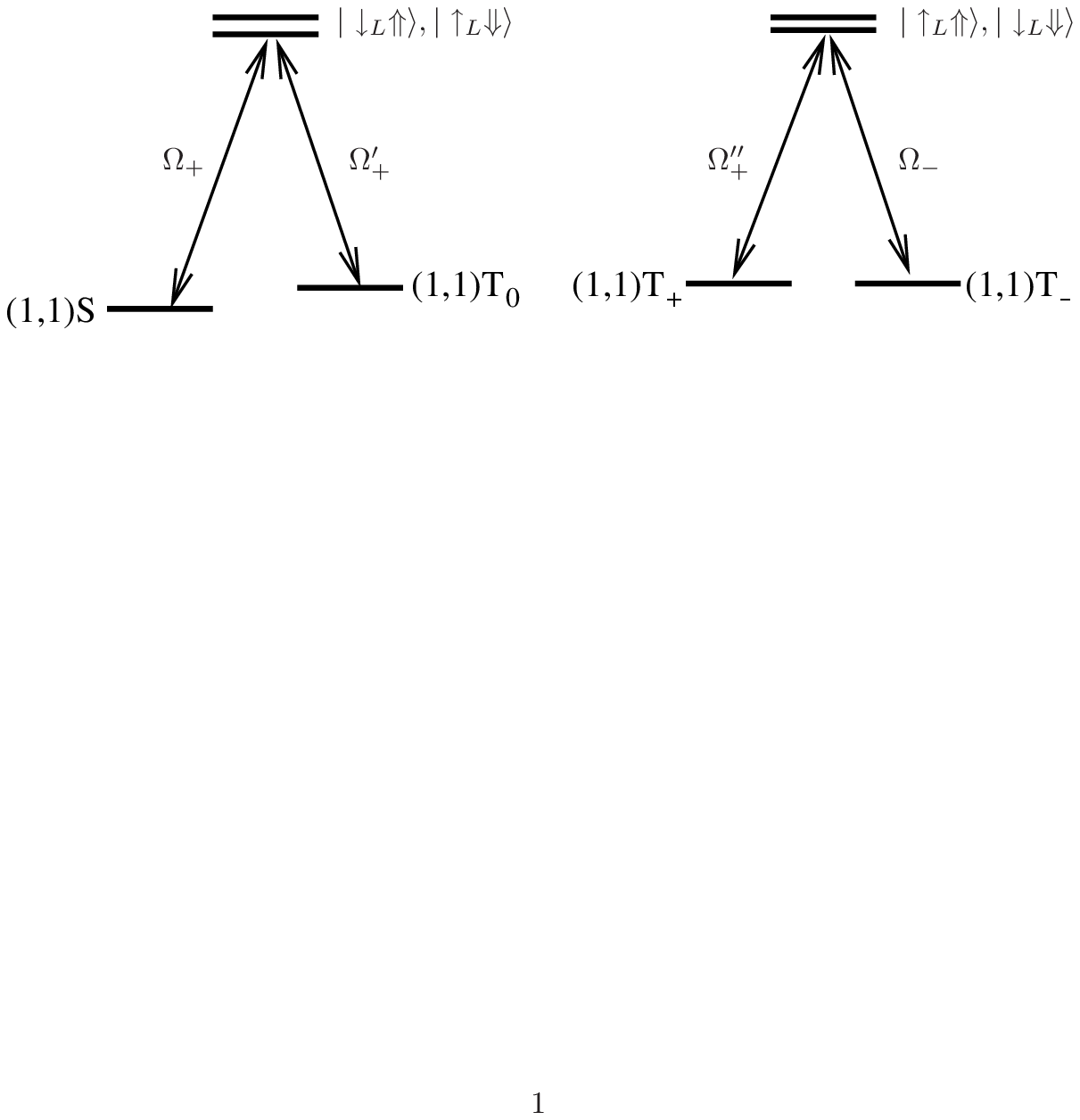}
\caption{Diagram showing the optical selection rules from the singlet-triplet ground state of the CQD to the intermediate excited states in a Raman spin-flip scheme. While $\bra{(1,1)S}$ and $\bra{(1,1)T_0}$ couple to the $|M|=1$ excited state manifold, $\bra{(1,1)T_+}$ and $\bra{(1,1)T_-}$ couple to the $|M|=2$ subspace. Here $M$ refers to the spin projection of the total pseudo-spin of the left electron and the right hole.}
\label{figOPTsel}
\end{figure}

Ideally, one would like to be able to connect all the members of the
$(1,1)$singlet-triplet space optically. The above discussion shows
however that $(T_+,T_-)$ and $(S,T_0)$ access different and
non-overlapping subspaces of the excited state manifold via optical
transitions for circular polarization. One can envision manipulating
the system with a combination of a static external magnetic field
and optical fields to overcome this problem.

\begin{figure}[hbt]
\includegraphics[clip,width=0.8\linewidth]{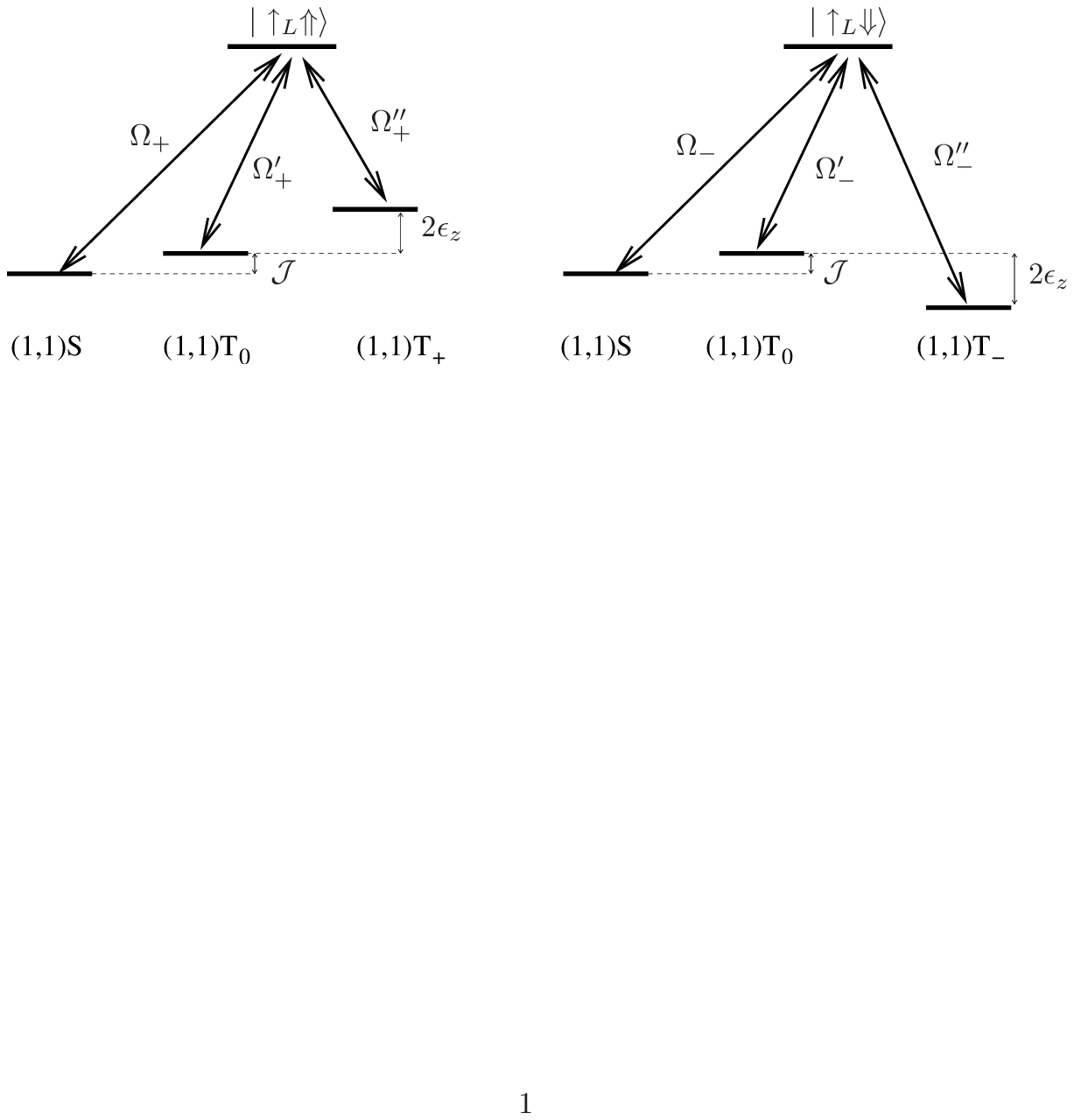}
\caption{Diagram showing the selection rules in the presence of an
in-plane magnetic field. The previously derived selection rules for
$B=0$ are no longer valid in this case, allowing for either $\bra{T_+}$ or
$\bra{T_-}$ state to be coupled to the same intermediate state as that of
$\bra{T_0}$ and $\bra{S}$. The two sets of ground state manifolds can be
accessed by using either left or right hand circularly polarized
fields.} \label{figQDselmix}
\end{figure}

Consider applying an in-plane magnetic field $\bm{B}=B\hat{\bm{x}}$.
This will mix both the ground state and the excited state manifolds.
Noting that the in plane hole $g$-factor is negligible, instead of
rewriting the new states in the $z$-representation, we will just
rewrite the interaction Hamiltonian in the new electron spin basis
\begin{equation}
V_+ = \frac{i\hbar}{2} \Omega_+ \left( \mcal{M}_{RR} (\erud-\erdd)\hrud + \mcal{M}_{LR} (\elud-\eldd)\hrud \right) \ex{-i\omega_+t} + h.c.
\end{equation}
Here, $\elud$ now creates an electron with $S_x^e=+1/2$
($\bra{\uparrow_L}$). The strong transitions now take the form
\begin{align*}
& V_+\bra{S} = -\mcal{M}_{RR} ( \bra{\uparrow_L\Uparrow} - \bra{\downarrow_L\Uparrow} ) \\
& V_+\bra{T_0} = -\mcal{M}_{RR}( \bra{\uparrow_L\Uparrow} + \bra{\downarrow_L\Uparrow} ) \\
& V_+\bra{T_+} =  -\mcal{M}_{RR} \bra{\uparrow_L\Uparrow} \\
& V_+\bra{T_-} = 0 \\
& V_-\bra{T_-} = -\mcal{M}_{RR} \bra{\uparrow_L\Downarrow}
\end{align*}
Thus, one can couple either the submanifold ($S$,$T_0$,$T_+$) by $\sigma_+$ light to $\bra{\uparrow_L\Uparrow}$ or ($S$,$T_0$,$T_-$)
by $\sigma_-$ to $\bra{\uparrow_L\Downarrow}$. The excitation scheme
for the circular polarization case is shown in
Fig.~\ref{figQDselmix}.

Note that, $\sigma_+$ polarized light couples $\bra{(1,1)S}$ and
$\bra{(1,1)T_0}$ to $\bra{\downarrow_L\Uparrow}$ as well. In this regard,
it is important to have a Zeeman splitting $\eps_Z
\lesssim \mcal{J}$ ($\eps_Z=g^e \mu_B B$ is the single electron Zeeman splitting) that is sufficiently large so that coupling of
a single spin state to a single intermediate excited optical state
may be possible.

Having an auxiliary state at disposal for optical manipulation is
important in the implementation of robust qubit rotations based on
stimulated Raman adiabatic passage (STIRAP) \cite{KisR02}. The
qubits in our CQD scheme above are formed by $(1,1)S$ and
$(1,1)T_0$, which is a subspace that can be protected from
hyper-fine induced dephasing by spin-echo techniques
\cite{PettaJTLYLMHG05}. Combined with the immunity to dephasing of
the intermediate state that can be achieved by STIRAP, the above
described scheme seems to be well suited for generation and manipulation of qubit states for quantum information protocols in Zinc-Blende (III-V)
semiconductor quantum dots.

\section{Optical mapping of spin states}

In this section we discuss a scheme to efficiently
prepare, manipulate and map spin states of a doubly charged CQD into
photon polarization for long distance quantum state transfer
\cite{CiracZKM97}.

Consider the CQD system placed inside a high-Q cavity and the gate
voltage tuned such that the system is in regime (II), but close to
the anti-crossing at $V=V_1^*$. We furthermore apply an in-plane
magnetic field. This will ensure that the states $S$,  $T_0$ and
$T_\pm$ are well-separated in energy. Let us assume that the system
is initialized to the superposition state \be \bra{\Psi}=\alpha
\bra{S} + \beta \bra{T_0} \label{eqinitSpin} \ee in the qubit space
composed of $(\bra{S},\bra{T_0})$. We first map this state into the
state
\begin{equation}
\bra{\Psi}=\alpha \bra{T_-} + \beta \bra{T_0}
\end{equation}
by a STIRAP sequence using $\sigma_-$ polarized light as decribed
for example in Ref.~\cite{renzoniS01}. The Raman nature of the
scattering will ensure a robust state mapping.

\begin{figure}[hbt]
\includegraphics[clip,width=0.4\linewidth]{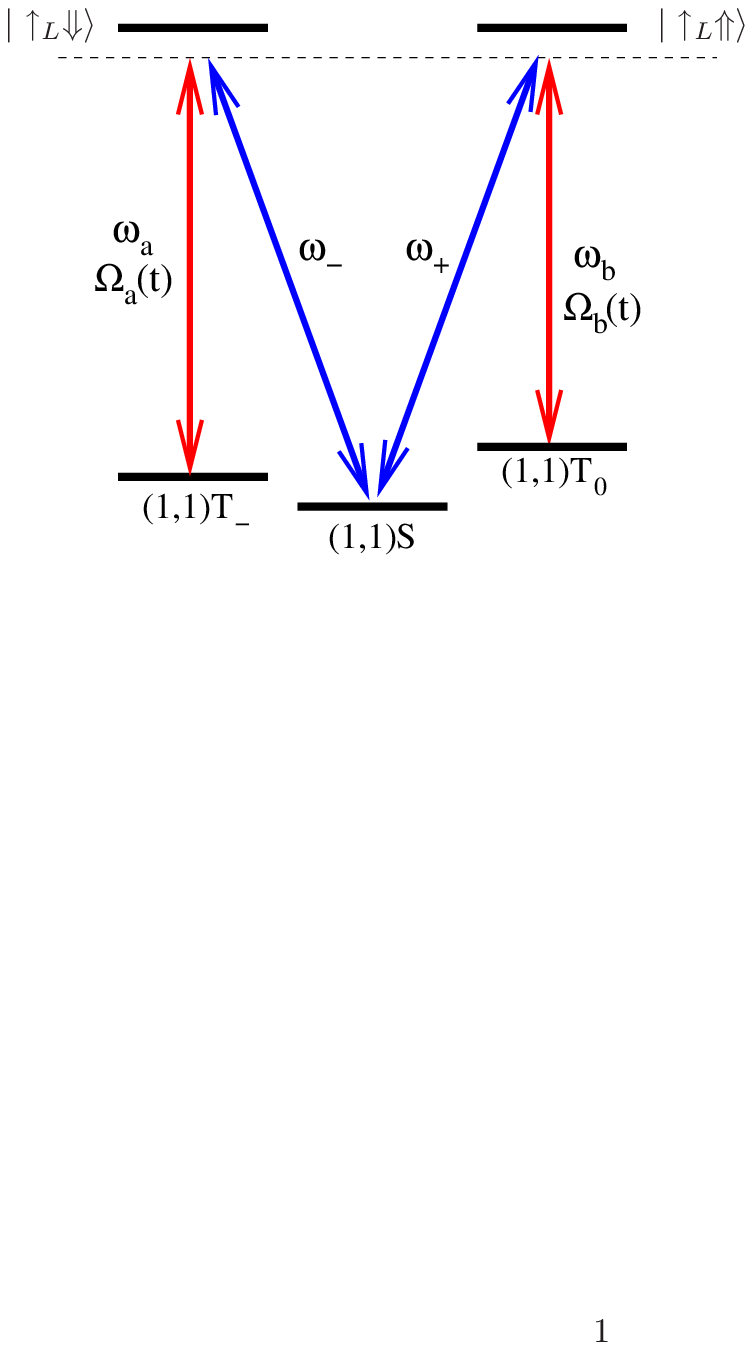}
\caption{The cavity assisted Raman scheme for spin-to-polarization
mapping. The $\sigma_-$ polarized laser at frequency $\om_a$ and
pulse shape $\Omega_a(t)$ results in cavity-assisted Raman
transition to $\bra{S}$ when the two-photon resonance condition $\om_- -
\om_a= E_{T_-} - E_S=\delta_1$ is satisfied. Similarly, the
$\sigma_+$ polarized laser at frequency $\om_b$ and pulse shape
$\Omega_b(t)$ results in cavity-assisted Raman transition to $\bra{S}$
when $\om_+ - \om_b=E_{T_0}-E_{S}=\delta_2$. Note that the energy splitting
between the excited states is negligible due to the small in-plane hole $g$-factors.} 
\label{figramanschem}
\end{figure}

We next turn on lasers with frequencies $\om_a$, $\om_b$,
polarizations $\sigma_-$, $\sigma_+$ and time dependent Rabi
frequencies $\Om_a(t)$ and $\Om_b(t)$, which are detuned by
$\delta_1$ and $\delta_2$ from $\bra{T_-} -
\bra{\uparrow_L\Downarrow}$ and  $\bra{T_0} -
\bra{\uparrow_L\Uparrow}$ transitions respectively. The detunings
$\delta_1$ and $\delta_2$ are carefully chosen so that the
transitions $\bra{\uparrow_L\Downarrow} - \bra{S}$ and
$\bra{\uparrow_L\Uparrow} - \bra{S}$ are resonant with a pair of
degenerate cavity modes $\om_-$ and $\om_+$ with orthogonal
polarizations (see Fig.~\ref{figramanschem}). Such cavities can be
engineered using photonic band gap structures where the cavity-mode
splitting can be precisely tuned by AFM oxidation so as to enforce
the degeneracy condition required \cite{HennessyHHBI06}. The
two-photon resonance condition for both polarizations will result in
cavity assisted Raman transition down to $\bra{S}$ and a transfer of the
original spin state (\ref{eqinitSpin}) to the polarization of the
emitted single photon via
\begin{equation}
( \alpha \bra{T_-} + \beta \bra{T_0} )\bra{0}_{ph} \rightarrow
\bra{S} ( \alpha \bra{1_{\sigma_-}}_{ph} + \beta
\bra{1_{\sigma_+}}_{ph} )
\end{equation}
The resulting photons can then be used for long-range quantum
communication in a distributed network of nodes containing
CQD-cavity systems. For this to work reliably, we require the Purcell-enhanced decay rate to be faster than spin dephasing rates. Experimental parameters \cite{BadolatoHADHPI05} suggest this can be achieved by a factor of 100.

\section{Coherent optical manipulation of coupled quantum dots with quasi-degenerate valence bands}

We now show that coherent optical coupling between fine structure
states of doubly-charged double quantum dots could be achieved even
when the intrinsic spin-orbit coupling of the semiconductor material
is weak.  We will rely upon an external magnetic field and weak
electron-hole symmetry breaking (differing electron and hole
$g$-factors) to show that all relevant Raman transitions are
accessible.

Consider the possibility of optical manipulation of quantum dots
where the spin-orbit effects are weak. The weakened selection rules
result in the following optical interaction Hamiltonian \be V_+ =
\frac{i\hbar}{2} \Omega_+ \mcal{M}_{RR} \left( \hd{R,3/2,3/2}
\ed{R,1/2,-1/2} + \frac{1}{\sqrt{3}} \, \hd{R,3/2,1/2}
\ed{R,1/2,1/2} - \sqrt{\frac{2}{3}}\, \hd{R,1/2,1/2} \ed{R,1/2,1/2}
\right) \ex{-i\omega_+t} + h.c. \ee Here, $\hd{R,J,J_z}$ creates a
hole on the right dot with total angular momentum
$\bm{J}^h=\bm{L}^h+\bm{S}^h$ and magnetic quantum number $J_z$,
while $\ed{R,1/2,\sigma}$ creates an electron in the conduction band
with spin projection $\sigma = \pm 1/2$. We only keep the strong,
direct transition terms of the optical Hamiltonian. Considering now
the action of $\sigma_+$ optical excitation on the $(1,1)$ ground
state manifold, we obtain
\begin{align*}
& V_+\bra{S} = \mcal{M}_{RR} ( \bra{(1\downarrow,2)\,3/2,3/2} + \frac{1}{\sqrt{3}}\, \bra{(1\uparrow,2)\,3/2,1/2} -   \sqrt{\frac{2}{3}}\, \bra{(1\uparrow,2)\,1/2,1/2} ) \\
& V_+\bra{T_0} =  \mcal{M}_{RR} ( -\bra{(1\downarrow,2)\,3/2,3/2} + \frac{1}{\sqrt{3}}\, \bra{(1\uparrow,2)\,3/2,1/2} -   \sqrt{\frac{2}{3}}\, \bra{(1\uparrow,2)\,1/2,1/2} ) \\
& V_+\bra{T_+} = - \mcal{M}_{RR} \bra{(1\uparrow,2)\,3/2,3/2}   \\
& V_+\bra{T_-} = \mcal{M}_{RR} ( \frac{1}{\sqrt{3}}\, \bra{(1\downarrow,2)\,3/2,1/2} - \sqrt{\frac{2}{3}}\,\bra{(1\downarrow,2)\,1/2,1/2} )
\end{align*}

\begin{figure}[hbt]
\includegraphics[clip,width=0.2\linewidth]{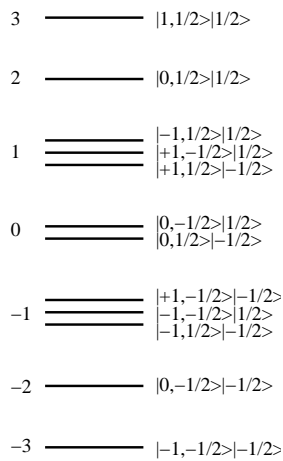}
\caption{The energy level structure of the excited state manifold.
On the left, the energies are given in units of $\mu_B B_x$. The
additional smaller splittings are due to the possibly different
$g$-factors of electrons and holes, i.e. nonzero $\eps^{h,e}$. The
states are written out explicitly in the format
$\bra{L_x^h,S_x^h}\bra{S_x^e}$ where the electron spin refers to the
left electron and the hole is on the right. The doubly occupied
singlet electrons in the right quantum dot are not shown.}
\label{figmultiplet}
\end{figure}

In principle, the degeneracy in the excited state manifold presents
a problem for coherent optical protocols due to potential
destructive quantum interference between different pathways. To
overcome this difficulty, we consider the effect of a magnetic
field. For simplicity we assume a Voigt configuration ($\bm{B}=
B\hat{\bm{x}}$, say). Writing the Zeeman Hamiltonian as
\begin{equation}
H_Z = \mu_B \, B_x \cdot ( L_x^{h} + g^hS_x^h + g^e S_x^e )
\end{equation}
where $L_x^{h}$ and $S_x^h$ are the $x$-components of the hole
orbital and spin angular momentum respectively. Let us take
$g^{h,e}=2+\eps^{h,e}$ with the hindsight that the spin-orbit
interaction is weak. Then, one can easily see that the total
angular momentum is not conserved. Thus we have to use the $(L^h,L^h_x,S^h,S^h_x,S^e_x)$ basis. The energy level
structure is given in Fig.~\ref{figmultiplet}. In this basis, the
optical interaction Hamiltonian is given by \be V_+ =
\frac{i\hbar}{2} \Omega_+ \mcal{M}_{RR} \left(
H_{\downarrow}^{\dagger} \, \ed{R\uparrow} +
H_{\uparrow}^{\dagger} \, \ed{R\downarrow} \right) + h.c. \ee
where
\begin{equation}
H_\sigma^{\dagger} = \frac{1}{2} \hd{R,-1,\sigma} - \frac{1}{\sqrt{2}} \hd{R,0,\sigma} -  \frac{1}{2} \hd{R,+1,\sigma} \quad \sigma = \uparrow \, , \, \downarrow
\end{equation}
The optical ground state is still given as above with the magnetic
quantum numbers measured along $x$ and with additional Zeeman energy
contributions. The Zeeman contributions also split up the excited
state manifold as shown in Fig.~\ref{figmultiplet}. Consider for
instance the transitions to the final state $\bra{f} =
\bra{-1,\downarrow}_h\bra{\uparrow}_e = \ed{L\uparrow}
\ed{R\uparrow}\ed{R\downarrow} \hd{R,-1,\downarrow} $. We get
$V_+\bra{S,T_0} =V_+ (\ed{L\uparrow}\ed{R\downarrow} \pm
\ed{L\downarrow}\ed{R\uparrow} ) = \ed{R\uparrow}\ed{R\downarrow} (
\ed{L\uparrow} H_{\downarrow}^{\dagger} \mp  \ed{L\downarrow}
H_{\uparrow}^{\dagger} )$ which imply
\begin{align*}
& \ket{f} V_+ \bra{S}  = 1/2 \\
& \ket{f} V_+ \bra{T_0}  =  1/2 \\
& \ket{f} V_+ \bra{T_+}  =  0 \\
& \ket{f} V_+ \bra{T_-}  =  0
\end{align*}
However, if one is to consider a spin-flip Raman scattering from $S$
to $T_0$ there are two available paths via intermediate states
$\bra{f_1} = \bra{-1,\downarrow}_h\bra{\uparrow}_e$ and $\bra{f_2} =
\bra{-1,\uparrow}_h\bra{\downarrow}_e$. Only when the electron and
hole $g$-factors are different would these contributions allow a
non-zero transition amplitude, otherwise  the two available paths
will interfere destructively. Hence it should in principle be
possible to couple the states $S,T_0$ to each other via two-photon
processes if the electron and hole $g$-factors are different. In
practice, the $g$-factors should be sufficiently different to ensure
$|\mu_B \, B_x  (g^h - g^e)| > \gamma$, where $\gamma$ is the
broadening of the optically-excited states.

\section{Conclusion}

We have presented a scheme for optical manipulation of the
metastable singlet-triplet subspace of a doubly-charged CQD. We find
that with strong as well as weak spin-orbit interactions it is
possible to implement Raman spin-flip transitions between a singlet
and a triplet fine-structure-split ground state. Such Raman
transitions enable the implementation of arbitrary coherent
rotations robustly via quantum optical techniques previously
proposed for single atoms or ions.

A key issue is the identification of the conditions that need to be
satisfied in order to generate a spin-flip Raman scattering within a
singlet-triplet space. In the absence of spin-orbit coupling and a
degenerate p-like valence band, the optical Hamiltonian is a spin-0
operator. Thus, unless the spin-conservation law is broken in the
intermediate state, it will not be possible to flip $S$ to $T_0$. In
the strong spin orbit case, the optical interaction Hamiltonian
however is a reducible spin operator with non-zero projections on
spin-1 and spin-0 subspaces (where spin is actually a pseudo-spin
due to the restriction to the heavy-hole band). This particular
property makes it possible to optically connect $S$ and $T_0$.

We expect that our findings will stimulate experimental research
aimed at combining electrical and optical manipulation of confined
spin states. The fact that optical manipulation is possible even for
quantum dot structures with weak spin-orbit interaction enhances the
prospects for pursuing experimental realization of quantum
communication protocols in such systems.

\begin{acknowledgments}
HET would like to thank Mete Atat\"ure, Jan Dreiser and Alex H\"ogele for useful discussions.
\end{acknowledgments}


\begin{thebibliography}{18}
\expandafter\ifx\csname natexlab\endcsname\relax\def\natexlab#1{#1}\fi
\expandafter\ifx\csname bibnamefont\endcsname\relax
  \def\bibnamefont#1{#1}\fi
\expandafter\ifx\csname bibfnamefont\endcsname\relax
  \def\bibfnamefont#1{#1}\fi
\expandafter\ifx\csname citenamefont\endcsname\relax
  \def\citenamefont#1{#1}\fi
\expandafter\ifx\csname url\endcsname\relax
  \def\url#1{\texttt{#1}}\fi
\expandafter\ifx\csname urlprefix\endcsname\relax\def\urlprefix{URL }\fi
\providecommand{\bibinfo}[2]{#2}
\providecommand{\eprint}[2][]{\url{#2}}

\bibitem[{\citenamefont{Awschalom et~al.}(2002)\citenamefont{Awschalom,
  Samarth, and Loss}}]{awschalom01}
\bibinfo{editor}{\bibfnamefont{D.~D.} \bibnamefont{Awschalom}},
  \bibinfo{editor}{\bibfnamefont{N.}~\bibnamefont{Samarth}}, \bibnamefont{and}
  \bibinfo{editor}{\bibfnamefont{D.}~\bibnamefont{Loss}}, eds.,
  \emph{\bibinfo{title}{Semiconductor Spintronics and Quantum Computation}}
  (\bibinfo{publisher}{Springer-Verlag}, \bibinfo{address}{Berlin},
  \bibinfo{year}{2002}).

\bibitem[{\citenamefont{Michler et~al.}(2000)\citenamefont{Michler, Kiraz,
  Becher, Schoenfeld, Petroff, Zhang, Hu, and Imamoglu}}]{MichlerKBSPZHI00}
\bibinfo{author}{\bibfnamefont{P.}~\bibnamefont{Michler}},
  \bibinfo{author}{\bibfnamefont{A.}~\bibnamefont{Kiraz}},
  \bibinfo{author}{\bibfnamefont{C.}~\bibnamefont{Becher}},
  \bibinfo{author}{\bibfnamefont{W.~V.} \bibnamefont{Schoenfeld}},
  \bibinfo{author}{\bibfnamefont{P.~M.} \bibnamefont{Petroff}},
  \bibinfo{author}{\bibfnamefont{L.~D.} \bibnamefont{Zhang}},
  \bibinfo{author}{\bibfnamefont{E.}~\bibnamefont{Hu}}, \bibnamefont{and}
  \bibinfo{author}{\bibfnamefont{A.}~\bibnamefont{Imamoglu}},
  \bibinfo{journal}{Science} \textbf{\bibinfo{volume}{290}},
  \bibinfo{pages}{2282} (\bibinfo{year}{2000}).

\bibitem[{\citenamefont{Pelton et~al.}(2002)\citenamefont{Pelton, Santori,
  Vuckovic, Zhang, Solomon, Plant, and Yamamoto}}]{PeltonSVZSPY02}
\bibinfo{author}{\bibfnamefont{M.}~\bibnamefont{Pelton}},
  \bibinfo{author}{\bibfnamefont{C.}~\bibnamefont{Santori}},
  \bibinfo{author}{\bibfnamefont{J.}~\bibnamefont{Vuckovic}},
  \bibinfo{author}{\bibfnamefont{B.~Y.} \bibnamefont{Zhang}},
  \bibinfo{author}{\bibfnamefont{G.~S.} \bibnamefont{Solomon}},
  \bibinfo{author}{\bibfnamefont{J.}~\bibnamefont{Plant}}, \bibnamefont{and}
  \bibinfo{author}{\bibfnamefont{Y.}~\bibnamefont{Yamamoto}},
  \bibinfo{journal}{Physical Review Letters} \textbf{\bibinfo{volume}{89}}
  (\bibinfo{year}{2002}).

\bibitem[{\citenamefont{Imamoglu}(2000)}]{imamoglu00}
\bibinfo{author}{\bibfnamefont{A.}~\bibnamefont{Imamoglu}},
  \bibinfo{journal}{Fortschr. Phys.} \textbf{\bibinfo{volume}{48}},
  \bibinfo{pages}{987} (\bibinfo{year}{2000}).

\bibitem[{\citenamefont{Haljan et~al.}(2005)\citenamefont{Haljan, Lee,
  Brickman, Acton, Deslauriers, and Monroe}}]{haljan05}
\bibinfo{author}{\bibfnamefont{P.~C.} \bibnamefont{Haljan}},
  \bibinfo{author}{\bibfnamefont{P.~J.} \bibnamefont{Lee}},
  \bibinfo{author}{\bibfnamefont{K.-A.} \bibnamefont{Brickman}},
  \bibinfo{author}{\bibfnamefont{M.}~\bibnamefont{Acton}},
  \bibinfo{author}{\bibfnamefont{L.}~\bibnamefont{Deslauriers}},
  \bibnamefont{and} \bibinfo{author}{\bibfnamefont{C.}~\bibnamefont{Monroe}},
  \bibinfo{journal}{Physical Review A} \textbf{\bibinfo{volume}{72}},
  \bibinfo{eid}{062316} (\bibinfo{year}{2005}).

\bibitem[{\citenamefont{Krenner et~al.}(2005)\citenamefont{Krenner, Sabathil,
  Clark, Kress, Schuh, Bichler, Abstreiter, and Finley}}]{krenner05}
\bibinfo{author}{\bibfnamefont{H.~J.} \bibnamefont{Krenner}},
  \bibinfo{author}{\bibfnamefont{M.}~\bibnamefont{Sabathil}},
  \bibinfo{author}{\bibfnamefont{E.~C.} \bibnamefont{Clark}},
  \bibinfo{author}{\bibfnamefont{A.}~\bibnamefont{Kress}},
  \bibinfo{author}{\bibfnamefont{D.}~\bibnamefont{Schuh}},
  \bibinfo{author}{\bibfnamefont{M.}~\bibnamefont{Bichler}},
  \bibinfo{author}{\bibfnamefont{G.}~\bibnamefont{Abstreiter}},
  \bibnamefont{and} \bibinfo{author}{\bibfnamefont{J.~J.}
  \bibnamefont{Finley}}, \bibinfo{journal}{Physical Review Letters}
  \textbf{\bibinfo{volume}{94}}, \bibinfo{eid}{057402} (\bibinfo{year}{2005}).

\bibitem[{\citenamefont{Stinaff et~al.}(2006)\citenamefont{Stinaff, Scheibner,
  Bracker, Ponomarev, Korenev, Ware, Doty, Reinecke, and
  Gammon}}]{StinaffSBPKWDRG06}
\bibinfo{author}{\bibfnamefont{E.~A.} \bibnamefont{Stinaff}},
  \bibinfo{author}{\bibfnamefont{M.}~\bibnamefont{Scheibner}},
  \bibinfo{author}{\bibfnamefont{A.~S.} \bibnamefont{Bracker}},
  \bibinfo{author}{\bibfnamefont{I.~V.} \bibnamefont{Ponomarev}},
  \bibinfo{author}{\bibfnamefont{V.~L.} \bibnamefont{Korenev}},
  \bibinfo{author}{\bibfnamefont{M.~E.} \bibnamefont{Ware}},
  \bibinfo{author}{\bibfnamefont{M.~F.} \bibnamefont{Doty}},
  \bibinfo{author}{\bibfnamefont{T.~L.} \bibnamefont{Reinecke}},
  \bibnamefont{and} \bibinfo{author}{\bibfnamefont{D.}~\bibnamefont{Gammon}},
  \bibinfo{journal}{Science} \textbf{\bibinfo{volume}{311}},
  \bibinfo{pages}{636} (\bibinfo{year}{2006}).

\bibitem[{\citenamefont{Zanardi and Rasetti}(1997)}]{zanardi97}
\bibinfo{author}{\bibfnamefont{P.}~\bibnamefont{Zanardi}} \bibnamefont{and}
  \bibinfo{author}{\bibfnamefont{M.}~\bibnamefont{Rasetti}},
  \bibinfo{journal}{Phys. Rev. Lett.} \textbf{\bibinfo{volume}{79}},
  \bibinfo{pages}{3306} (\bibinfo{year}{1997}).

\bibitem[{\citenamefont{Lidar et~al.}(1998)\citenamefont{Lidar, Chuang, and
  Whaley}}]{lidar98}
\bibinfo{author}{\bibfnamefont{D.~A.} \bibnamefont{Lidar}},
  \bibinfo{author}{\bibfnamefont{I.~L.} \bibnamefont{Chuang}},
  \bibnamefont{and} \bibinfo{author}{\bibfnamefont{K.~B.}
  \bibnamefont{Whaley}}, \bibinfo{journal}{Phys. Rev. Lett.}
  \textbf{\bibinfo{volume}{81}}, \bibinfo{pages}{2594} (\bibinfo{year}{1998}).

\bibitem[{\citenamefont{Duan and Guo}(1997)}]{duan97}
\bibinfo{author}{\bibfnamefont{L.-M.} \bibnamefont{Duan}} \bibnamefont{and}
  \bibinfo{author}{\bibfnamefont{G.-C.} \bibnamefont{Guo}},
  \bibinfo{journal}{Phys. Rev. Lett.} \textbf{\bibinfo{volume}{79}},
  \bibinfo{pages}{1953} (\bibinfo{year}{1997}).

\bibitem[{\citenamefont{Petta et~al.}(2005)\citenamefont{Petta, Johnson,
  Taylor, Laird, Yacoby, Lukin, Marcus, Hanson, and Gossard}}]{PettaJTLYLMHG05}
\bibinfo{author}{\bibfnamefont{J.~R.} \bibnamefont{Petta}},
  \bibinfo{author}{\bibfnamefont{A.~C.} \bibnamefont{Johnson}},
  \bibinfo{author}{\bibfnamefont{J.~M.} \bibnamefont{Taylor}},
  \bibinfo{author}{\bibfnamefont{E.~A.} \bibnamefont{Laird}},
  \bibinfo{author}{\bibfnamefont{A.}~\bibnamefont{Yacoby}},
  \bibinfo{author}{\bibfnamefont{M.~D.} \bibnamefont{Lukin}},
  \bibinfo{author}{\bibfnamefont{C.~M.} \bibnamefont{Marcus}},
  \bibinfo{author}{\bibfnamefont{M.~P.} \bibnamefont{Hanson}},
  \bibnamefont{and} \bibinfo{author}{\bibfnamefont{A.~C.}
  \bibnamefont{Gossard}}, \bibinfo{journal}{Science}
  \textbf{\bibinfo{volume}{309}}, \bibinfo{pages}{2180} (\bibinfo{year}{2005}).

\bibitem[{\citenamefont{Jarillo-Herrero
  et~al.}(2004)\citenamefont{Jarillo-Herrero, Sapmaz, Dekker, Kouwenhoven, and
  van~der Zant}}]{Jarillo-HerreroSDKv04}
\bibinfo{author}{\bibfnamefont{P.}~\bibnamefont{Jarillo-Herrero}},
  \bibinfo{author}{\bibfnamefont{S.}~\bibnamefont{Sapmaz}},
  \bibinfo{author}{\bibfnamefont{C.}~\bibnamefont{Dekker}},
  \bibinfo{author}{\bibfnamefont{L.~P.} \bibnamefont{Kouwenhoven}},
  \bibnamefont{and} \bibinfo{author}{\bibfnamefont{H.~S.~J.}
  \bibnamefont{van~der Zant}}, \bibinfo{journal}{Nature}
  \textbf{\bibinfo{volume}{429}}, \bibinfo{pages}{389} (\bibinfo{year}{2004}).

\bibitem[{\citenamefont{Bahder}(1990)}]{Bahder90}
\bibinfo{author}{\bibfnamefont{T.~B.} \bibnamefont{Bahder}},
  \bibinfo{journal}{Physical Review B} \textbf{\bibinfo{volume}{41}},
  \bibinfo{pages}{11992} (\bibinfo{year}{1990}).

\bibitem[{\citenamefont{Kis and Renzoni}(2002)}]{KisR02}
\bibinfo{author}{\bibfnamefont{Z.}~\bibnamefont{Kis}} \bibnamefont{and}
  \bibinfo{author}{\bibfnamefont{F.}~\bibnamefont{Renzoni}},
  \bibinfo{journal}{Physical Review A} \textbf{\bibinfo{volume}{65}}
  (\bibinfo{year}{2002}).

\bibitem[{\citenamefont{Cirac et~al.}(1997)\citenamefont{Cirac, Zoller, Kimble,
  and Mabuchi}}]{CiracZKM97}
\bibinfo{author}{\bibfnamefont{J.~I.} \bibnamefont{Cirac}},
  \bibinfo{author}{\bibfnamefont{P.}~\bibnamefont{Zoller}},
  \bibinfo{author}{\bibfnamefont{H.~J.} \bibnamefont{Kimble}},
  \bibnamefont{and} \bibinfo{author}{\bibfnamefont{H.}~\bibnamefont{Mabuchi}},
  \bibinfo{journal}{Physical Review Letters} \textbf{\bibinfo{volume}{78}},
  \bibinfo{pages}{3221} (\bibinfo{year}{1997}).

\bibitem[{\citenamefont{{Renzoni} and {Stenholm}}(2001)}]{renzoniS01}
\bibinfo{author}{\bibfnamefont{F.}~\bibnamefont{{Renzoni}}} \bibnamefont{and}
  \bibinfo{author}{\bibfnamefont{S.}~\bibnamefont{{Stenholm}}},
  \bibinfo{journal}{Optics Communications} \textbf{\bibinfo{volume}{189}},
  \bibinfo{pages}{69} (\bibinfo{year}{2001}).

\bibitem[{\citenamefont{Hennessy et~al.}(2006)\citenamefont{Hennessy, Hogele,
  Hu, Badolato, and Imamoglu}}]{HennessyHHBI06}
\bibinfo{author}{\bibfnamefont{K.}~\bibnamefont{Hennessy}},
  \bibinfo{author}{\bibfnamefont{C.}~\bibnamefont{Hogele}},
  \bibinfo{author}{\bibfnamefont{E.}~\bibnamefont{Hu}},
  \bibinfo{author}{\bibfnamefont{A.}~\bibnamefont{Badolato}}, \bibnamefont{and}
  \bibinfo{author}{\bibfnamefont{A.}~\bibnamefont{Imamoglu}},
  \bibinfo{journal}{Applied Physics Letters} \textbf{\bibinfo{volume}{89}}
  (\bibinfo{year}{2006}).

\bibitem[{\citenamefont{Badolato et~al.}(2005)\citenamefont{Badolato, Hennessy,
  Atature, Dreiser, Hu, Petroff, and Imamoglu}}]{BadolatoHADHPI05}
\bibinfo{author}{\bibfnamefont{A.}~\bibnamefont{Badolato}},
  \bibinfo{author}{\bibfnamefont{K.}~\bibnamefont{Hennessy}},
  \bibinfo{author}{\bibfnamefont{M.}~\bibnamefont{Atature}},
  \bibinfo{author}{\bibfnamefont{J.}~\bibnamefont{Dreiser}},
  \bibinfo{author}{\bibfnamefont{E.}~\bibnamefont{Hu}},
  \bibinfo{author}{\bibfnamefont{P.~M.} \bibnamefont{Petroff}},
  \bibnamefont{and} \bibinfo{author}{\bibfnamefont{A.}~\bibnamefont{Imamoglu}},
  \bibinfo{journal}{Science} \textbf{\bibinfo{volume}{308}},
  \bibinfo{pages}{1158} (\bibinfo{year}{2005}).

\end{thebibliography}
\end{document}